\documentclass[11pt]{article}

\title{Kundt Spacetimes as Solutions of Topologically Massive Gravity}

\usepackage{amsmath}
\usepackage{amsfonts}
\usepackage{amssymb}
\usepackage{amscd}
\usepackage{latexsym}
\usepackage{setspace}
\usepackage{rotating}

\spacing{1.4}

\makeatletter
\@addtoreset{equation}{section}
\makeatother

\newcommand{\be}{\begin{equation}}
\newcommand{\ee}{\end{equation}}
\newcommand{\ben}{\begin{equation}}
\newcommand{\een}{\end{equation}}
\newcommand{\bea}{\setlength\arraycolsep{2pt} \begin{eqnarray}}
\newcommand{\eea}{\end{eqnarray}}

\newcommand{\nnr}{\nonumber \\}
\newcommand{\la}{\label}

\newcommand{\bib}{\bibitem}
\newcommand{\ci}{\cite}
\newcommand{\se}{\section}
\newcommand{\sse}{\subsection}
\newcommand{\ssse}{\subsubsection}
\newcommand{\pa}{\paragraph}
\newcommand{\qd}{\quad}

\newcommand{\lt}{\left}
\newcommand{\rt}{\right}
\newcommand{\fr}{\frac}

\newcommand{\half}{\tfrac{1}{2}}
\newcommand{\tf}{\tfrac}

\newcommand{\wt}{\widetilde}

\newcommand{\cosec}{\textrm{cosec}}
\newcommand{\cosech}{\textrm{cosech}}

\newcommand{\leri}{\leftrightarrow}

\newcommand{\na}{\nabla}
\newcommand{\ov}{\overline}
\newcommand{\pd}{\partial}
\newcommand{\ra}{\rightarrow}
\newcommand{\sech}{\textrm{sech}}
\newcommand{\sr}{\sqrt}

\newcommand{\ga}{\alpha}

\newcommand{\gc}{\gamma}

\newcommand{\gd}{\delta}

\newcommand{\gep}{\epsilon}

\newcommand{\gy}{\eta}

\newcommand{\gq}{\theta}

\newcommand{\gL}{\Lambda}

\newcommand{\gr}{\rho}

\newcommand{\gs}{\sigma}

\newcommand{\gw}{\omega}

\newcommand{\ud}{\textrm{d}}
\newcommand{\ue}{\textrm{e}}

\newcommand{\ui}{\textrm{i}}

\newcommand{\us}{\textrm{s}}

\newcommand{\uE}{\textrm{E}}
\newcommand{\uL}{\textrm{L}}

\newcommand{\im}{\textrm{i}}

\newcommand{\bbR}{\mathbb{R}}

\newcommand{\tW}{\widetilde{W}}

\newcommand{\tlf}{\widetilde{f}}

\newcommand{\tlu}{\widetilde{u}}
\newcommand{\tlv}{\widetilde{v}}

\newcommand{\ds}{\textrm{d} s^2}

\begin{document}

\begin{titlepage}
\begin{flushright}
MIFP-09-49
\end{flushright}
\vspace*{100pt}
\begin{center}
{\bf \Large{Kundt Spacetimes as Solutions of Topologically Massive Gravity}}\\
\vspace{50pt}
\long\def\symbolfootnote[#1]#2{\begingroup
\def\thefootnote{\fnsymbol{footnote}}\footnote[#1]{#2}\endgroup}
\large{David D.K. Chow$^1$, C.N. Pope$^{1, 2}$ and Ergin Sezgin$^1$}
\end{center}

\begin{center}

$^1${\it George P. and Cynthia W. Mitchell Institute for
Fundamental Physics \& Astronomy,\\
Texas A\&M University, College Station, TX 77843-4242, USA}\\
\vspace*{10pt}
$^2${\it Department of Applied Mathematics and Theoretical Physics, University
of Cambridge,\\
Centre for Mathematical Sciences, Wilberforce Road, Cambridge CB3 0WA, UK}\\

\vspace{50pt}

{\bf Abstract}

\end{center}

\noindent We obtain new solutions of topologically massive gravity.  We find the general Kundt solutions, which in three dimensions are spacetimes admitting an expansion-free null geodesic congruence.  The solutions are generically of algebraic type II, but special cases are types III, N or D.  Those of type D are the known spacelike-squashed AdS$_3$ solutions, and of type N are the known AdS pp-waves or new solutions.  Those of types II and III are the first known solutions of these algebraic types.  We present explicitly the Kundt solutions that are CSI spacetimes, for which all scalar polynomial curvature invariants are constant, whereas for the general case we reduce the field equations to a series of ordinary differential equations.  The CSI solutions of types II and III are deformations of spacelike-squashed AdS$_3$ and the round AdS$_3$, respectively.

\end{titlepage}

\tableofcontents

\newpage


\se{Introduction}


Topologically massive gravity (TMG) \cite{Deser:1982vya, Deser:1982vyb, Deser:1982sv} is a three-dimensional gravitational theory described by the field equation
\ben
G_{\mu \nu} + \gL g_{\mu \nu} + \fr{1}{\mu} C_{\mu \nu} = 0 \, , \la{fieldeq2}
\een
where $C_{\mu \nu} = \gep{_\mu}{^{\gr \gs}} \na_\gr (R_{\gs \nu} - \tf{1}{4} R g_{\gs \nu})$ is the Cotton tensor.  We shall mostly take $\gL = - m^2$ to be negative, but sometimes take $m = 0$ or $\gL = m^2 > 0$.  Of particular recent interest \cite{LSS} has been the theory at the chiral point, which has $\mu = m = \sr{- \gL}$.  Like in Einstein gravity, there are trivial solutions that are maximally symmetric; these are characterized in TMG as being the only Einstein solutions, or equivalently the only solutions with a vanishing Cotton tensor.

In a previous paper \cite{chposeI}, the known exact solutions of TMG were reviewed.  Although there are many solutions in the literature, obtained by different approaches, it was found that many of these are locally the same, related by coordinate transformations.  In particular, in the presence of a non-vanishing cosmological constant, all non-trivial solutions in the literature are locally one of three ubiquitous solutions \cite{chposeI}:
\begin{enumerate}
\item Timelike-squashed AdS$_3$,
\item Spacelike-squashed AdS$_3$,
\item AdS pp-waves.
\end{enumerate}
It is surprising that so many different approaches should all lead the same three solutions.  In the absence of a cosmological constant, the only further solution known is a triaxial squashing of AdS$_3$, rather than just the biaxial squashing that are timelike- and spacelike-squashed AdS$_3$.  This motivates us to ask: are these the only solutions of TMG?  In this paper, we shall search for and obtain further solutions.

Before proceeding with this task, we should gain some understanding of why finding further solutions of TMG has been hard.  We recall a few results in the literature that place constraints on possible solutions, and so provide some reasons why so few solutions are known.

One usually expects the simplest exact solutions of a theory to be static, or more generally to admit a hypersurface-orthogonal Killing vector.  In the context of TMG, one might try to find such solutions.  However, it has been shown \ci{alinut, cavaglia, desfra} that, if the hypersurface-orthogonal Killing vector is non-null\footnote{\cite{alinut} clearly uses the non-null property, even though it is not explicitly stated.}, such solutions are Einstein.  In particular, there are no static solutions that are not Einstein.  This is a strong constraint, and is the main reason why few exact solutions of TMG are known.

However, there are non-Einstein solutions that are stationary but not static, such as the squashed AdS$_3$ solutions.  A weaker assumption involving Killing vectors would be to assume two commuting Killing vectors --- one timelike and one spacelike --- that are not hypersurface-orthogonal.  The problem of finding such solutions of TMG can be reduced to a particle motion problem in a three-dimensional Minkowski spacetime \ci{clement}.  Using this formalism, it has been shown that the only stationary and rotationally symmetric solutions of chiral gravity, which has certain asymptotic behaviour and a particular value of the gravitational Chern--Simons coupling constant, are locally AdS$_3$ \ci{masost}.

The three solutions that are ubiquitous in the literature can be characterized in a simple coordinate-invariant manner.  The timelike- and spacelike-squashed AdS$_3$ solutions are the only algebraic type D solutions \cite{chposeI}, and the AdS pp-waves are the only solutions that admit a null Killing vector \cite{gipose}.

Given these results, but anticipating that we shall find more solutions, we are motivated to ask: is TMG so constrained that it is feasible to find all exact solutions?  Less ambitiously, one could restrict to particular values of the coupling constant and to certain asymptotic behaviour, but in this paper we do not impose such restrictions and search for local metrics that solve TMG, regardless of their asymptotics.

As an initial step in finding and classifying further solutions, we shall take an approach that has been used in four-dimensional general relativity, where it gives some important classes of exact solutions.  The approach is to consider spacetimes that admit a ``special'' null vector field $k^\mu$.  For example, the requirement that a spacetime admits a null Killing vector field is a strong restriction: all such vacuum solutions of four-dimensional Einstein gravity (without a cosmological constant) are known \ci{dautcourt} (see also Chapter 24 of \ci{stkrmahohe}).  If the null Killing vector is covariantly constant, i.e.~$\na_\mu k_\nu = 0$, then the solutions are (by definition) pp-waves.

The pp-wave spacetimes are special cases of a larger class of spacetimes
admitting a special null vector field: the Kundt spacetimes.  These are
spacetimes admitting a null geodesic vector field that is expansion-free,
shear-free and twist-free (see, for example, Chapter 31 of \ci{stkrmahohe}).  A further complication would be to drop the expansion-free condition $\na_\mu k^\mu = 0$, so that we have a spacetime admitting a null geodesic vector field that is expanding, shear-free and twist-free; these are the Robinson--Trautman spacetimes (see, for example, Chapter 28 of \ci{stkrmahohe}).

Recently, there have been studies, in general (higher) dimensions, of Kundt spacetimes \cite{podzof, cohepape}, and of Robinson--Trautman spacetimes \cite{podort}.  In general dimensions, Kundt and Robinson--Trautman spacetimes are defined by the above conditions on the expansion, shear and twist of a null geodesic vector field.  However, in three dimensions a null geodesic vector field is automatically shear-free and twist-free.  Therefore the Kundt and Robinson--Trautman spacetimes have perhaps more significance in three dimensions, since they exhaust the Lorentzian geometries.

A common property of the existing solutions of TMG is that all scalars constructed as polynomials in the curvature and its derivatives are constant; such spacetimes are called constant scalar invariant (CSI) spacetimes \cite{cohepe1}.  All CSI spacetimes are known in three dimensions \ci{cohepe2} (they have also recently been investigated in four dimensions \ci{cohepe3}).  The initial classification is into two classes: locally homogeneous spacetimes, and Kundt spacetimes (or possibly both).  This is a further motivation for studying the Kundt spacetimes as solutions of TMG.

In this paper, we shall find the most general Kundt spacetime that solves TMG, finding several classes of new solutions.  By this, we mean that we can reduce the problem to solving a series of three ordinary differential equations, one of which is non-linear, whereas the other two are linear.  However, for the CSI solutions of TMG, we are able to present the metrics explicitly by solving the differential equations explicitly.  Of the existing solutions of TMG, the AdS pp-wave and spacelike squashed AdS$_3$ solutions are Kundt spacetimes, but timelike-squashed AdS$_3$ and the $\gL = 0$ theory's triaxially squashed AdS$_3$ solution are instead Robinson--Trautman.

We shall also examine the algebraic classification of the Kundt solutions of TMG.  The generic Kundt solution is of Petrov--Segre type II, but special cases are types III, D and N (and O).  These provide the first examples of solutions of TMG with Petrov--Segre types II and III.  Kundt solutions of Petrov--Segre types III, D and N (and O) are necessarily CSI; some solutions of Petrov--Segre type II are CSI.

The CSI Kundt solutions naturally split into two classes.  One CSI Kundt class describes deformations of spacelike-squashed AdS$_3$ and is generically type II; a special case is spacelike-squashed AdS$_3$, which is type D.  The other CSI Kundt class describes deformations of the round AdS$_3$, and which are generically of type III, and splits into a number of families; this number depends on the sign of $\gL$.  Each family contains a special case that is type N; if $\gL \leq 0$, then for one family, this type N solution is the (AdS) pp-wave, but the other type N solutions are new.

The outline of the rest of this paper is as follows.  In Section 2, we find the
most general Kundt solution of TMG.  We highlight as special cases those with
the further simplifying assumption that their scalar polynomial curvature
invariants are all constant.  In Sections 3 and 4 we present explicitly Kundt
solutions of TMG with this property: those in Section 3 are of Petrov--Segre
type II, corresponding to deformations of spacelike-squashed AdS$_3$; those in
Section 4 are of type III, corresponding to deformations of the round AdS$_3$.  We conclude in Section 5.


\se{Kundt Solutions of Topologically Massive Gravity}



\sse{Kundt spacetimes}


To define the Kundt spacetimes, we first review null geodesic congruences, focusing on 3 dimensions.  Suppose that $k^\mu$ is a null geodesic congruence, so $k^\mu \na_\mu k^\nu = 0$.  For null geodesic congruences in general dimensions $D \geq 3$, we define the following optical scalars:
\begin{enumerate}
\item Expansion: $\gq = \fr{1}{D-2} \na_\mu k^\mu$,
\item Shear: $\gs^2 = (\na^\mu k^\nu) \na_{( \mu} k_{\nu )} - \fr{1}{D-2}
(\na_\mu k^\mu)^2$,
\item Twist: $\gw^2 = (\pd^\mu k^\nu) \pd_{[ \mu} k_{\nu ]}$.
\end{enumerate}
An alternative approach to defining these scalars is to project $\na_\mu k_\nu$
onto a $(D-2)$-dimensional subspace by introducing an auxiliary null vector
field $l^\mu$ that satisfies $k^\mu l_\mu = -1$.  We define the projection
$h_{\mu \nu} = g_{\mu \nu} + k_\mu l_\nu + l_\mu k_\nu$, and $\hat{B}_{i j} =
h_i^\mu h_j^\nu \na_\mu k_\nu$, where $i$, $j$ are $(D-2)$-dimensional indices
for the subspace transverse to $k$ and $l$.  Then we decompose $\hat{B}_{ij}$ as
\ben
\hat{B}_{ij} = \fr{1}{D-2} \gq h_{ij} + \gs_{ij} + \gw_{ij} \, ,
\een
where the expansion is $\gq$, the shear (matrix) $\gs_{ij}$ is symmetric and
traceless, and the twist (matrix) $\gw_{ij}$ is antisymmetric.  Although
$\gs_{ij}$ and $\gw_{ij}$ depend on the choice of $l^\mu$, if they vanish for
one choice of $l^\mu$, then they vanish for all choices.  The (scalar) shear
$\gs^2 = \gs^{ij} \gs_{ij}$ and the (scalar) twist $\gw^2 = \gw^{ij} \gw_{ij}$
are independent of the choice of $l^\mu$. 

If the spacetime is 3-dimensional, then the projection is onto a 1-dimensional
subspace, and so we have
\ben
m_\mu m_\nu = g_{\mu \nu} + k_\mu l_\nu + l_\mu k_\nu \, ,
\een
where $m^\mu m_\mu = 1$ and $k^\mu m_\mu = l^\mu m_\mu = 0$.  Then we can expand
$\ud k = \pd_\mu k_\nu \,  \ud x^\mu \,  \wedge \ud x^\nu$ in terms of the basis
$( k \wedge l, k \wedge m, l \wedge m )$.  Since $k$ is null and geodesic, we
have $k^\mu \pd_{[\mu} k_{\nu]} = 0$.  Therefore $\ud k$ is proportional (up to
a scalar function) to $k \wedge m$, or equivalently $* \ud k$ is proportional to
$l$.  Hence $k$ must be hypersurface-orthogonal, i.e.~$k \wedge \ud k = 0$. 
From the 3-dimensional perspective, the twist vanishes because $* \ud k$ is
proportional to $l$, which is null, and the shear vanishes as a consequence of
$k^\mu \na_{(\mu} k_{\nu)} = 0$ implying that $\na_{(\mu} k_{\nu )}$ is a linear
combination (with scalar function coefficients) of $k_\mu k_\nu$, $k_{(\mu}
m_{\nu)}$ and $m_\mu m_\nu$.  A shorter argument is that $\hat{B}_{ij}$ is
defined on a 1-dimensional subspace, so the shear $\gs_{ij}$ and the twist
$\gw_{ij}$ trivially vanish, or equivalently $\gs^2$ and $\gw^2$ vanish.  In
summary, every null geodesic congruence of a three-dimensional spacetime is
shear-free and twist-free.

The defining property of a $D$-dimensional Kundt spacetime is a metric admitting a null geodesic vector field $k^\mu$ that is:
\begin{enumerate}

\item Expansion-free: \,  $\na_\mu k^\mu = 0$,

\item Shear-free: \, $(\na^\mu k^\nu) \na_{(\mu} k_{\nu)} - \fr{1}{D-2} (\na_\mu
k^\mu)^2 = 0$,

\item Twist-free: \, $(\pd^\mu k^\nu) \pd_{[\mu} k_{\nu]} = 0$.

\end{enumerate}
In three dimensions, a Kundt spacetime is simply one that admits an expansion-free null geodesic congruence.  Similarly, a three-dimensional spacetime that admits an expanding null geodesic congruence, i.e.~with $\na_\mu k^\mu \neq 0$, is a Robinson--Trautman spacetime.

Turning to TMG, we shall see that the AdS pp-wave solution and the biaxial spacelike-squashed AdS$_3$ solution are Kundt, whilst, by elimination, the biaxial timelike-squashed AdS$_3$
solution and the $\gL = 0$ theory's triaxially squashed AdS$_3$ solutions are
Robinson--Trautman.  In contrast with non-trivial solutions of TMG that admit a
non-null Killing vector, which cannot be hypersurface-orthogonal, if a solution
of TMG admits a null Killing vector, which must be geodesic, then it is
hypersurface-orthogonal.  The AdS pp-wave arises as the general supersymmetric
solution of topologically massive supergravity, the supersymmetric extension of
TMG, and are the general solutions of TMG that admit a null Killing vector
\cite{gipose}.  Supersymmetry means that a solution possesses a Killing spinor
$\gep$, which in turn induces a null Killing vector field $k^\mu = \ov{\gep}
\gc^\mu \gep$.  Perhaps surprisingly, in topologically massive supergravity the
converse is true: a spacetime that admits a null Killing vector is
supersymmetric (provided the orientation is correctly chosen).  All such
spacetimes can be written in generalized Kerr--Schild form,
\ben
g_{\mu \nu} = \gy_{\mu \nu} + f k_\mu k_\nu ,
\een
where $\gy_{\mu \nu}$ is a ``background'' metric that is AdS$_3$, $k_\mu \,  \ud
x^\mu$ is a 1-form that is null with respect to both the AdS$_3$ background
metric and the full metric, and $f$ is a scalar function.


\sse{General Kundt solutions}


We now seek some new local metrics that give solutions of TMG by considering Kundt spacetimes.  Despite TMG being a higher-derivative theory, the field equation is still linear in curvature, and so the calculations and results bear some similarity to those for Kundt solutions of four-dimensional Einstein
gravity.

A general three-dimensional Kundt spacetime can be written in
the form
\ben
\ds = \fr{\ud \gr^2}{P(u, \gr)^2} + 2 \,  \ud u \,  \ud v +
f(v, u, \gr) \,  \ud u^2 + 2 W(v, u, \gr) \,  \ud u \,  \ud \gr \, .
\la{Kundtmetric}
\een
By the coordinate transformation $\gr' = \int \ud \gr / P(u, \gr)$,
which maintains the general form of the metric, we henceforth take $P(u, \gr) =
1$.  Our orientation convention is $\gep_{v u \gr} = 1$.  The null vector in the definition of a Kundt spacetime is $k^\mu \pd_\mu = \pd / \pd v$, which has
associated 1-form $k_\mu \,  \ud x^\mu = \ud u$.  We shall use $\,  \dot{} \, $ to denote $\pd / \pd u$, and $'$ to denote $\pd / \pd \gr$.

Consider a Kundt spacetime that is a solution of TMG.  The $vv$ component of the
field equation gives $\pd^3 W / \pd v^3 = 0$, and hence $W$ is of the form
\ben
W(v, u, \gr) = v^2 W_2 (u, \gr) + v W_1 (u, \gr) + W_0 (u, \gr) \, .
\een
Then the $v \gr$ component of the field equation gives $\pd^3 f / \pd v^3 = 4
\mu W_2 (u, \gr)$.  Therefore $f$ is of the form
\ben
f(v, u, \gr) = v^3 f_3 (u, \gr) + v^2 f_2 (u, \gr) + v f_1 (u, \gr) 
+ f_0 (u, \gr) \, ,
\een
and so $W_2 (u, \gr) = \tf{3}{2} f_3 (u, \gr) / \mu$.  Then the 
Ricci scalar is
\ben
R = - \fr{45 v^2}{2 \mu^2} f_3^2 
+ \fr{3 v}{\mu} (2 f_3' + 2 \mu f_3 - 5 W_1 f_3) + 2 f_2 
+ 2 W_1' - \fr{3}{2} W_1^2 - \fr{6 W_0 f_3}{\mu} \, .
\een
Since a solution of TMG has a constant Ricci scalar $R = 6 \gL$, 
we therefore have $f_3 = 0$, and so $W_2 = 0$.  In fact, more generally, for
three-dimensional Kundt spacetimes solving $G_{\mu \nu} = 8 \pi T_{\mu \nu}$,
with $T_{v \gr} = 0$ and $\pd_v T{^\mu}{_\mu} = 0$, $f$ must be quadratic in $v$
and $W$ must be linear in $v$, which is a special case of an analogous result in
general dimensions \cite{podzof}.  It follows that
\ben
f_2 (u, \gr) = - W_1' + \fr{3}{4} W_1^2 + 3 \gL \, .
\la{f2Kundt}
\een

It is useful at this point to give the coordinate transformations that leave the
general form of such a Kundt metric invariant.  This form of the metric is
preserved by the coordinate changes (see, for example, \ci{podzof})
\ben
v = \fr{\tlv}{\dot{u} (\tlu)} + F(\wt{u}, \wt{\gr}),
\qd u = u(\tlu) , \qd \gr = \wt{\gr} + G(\tlu) \, ,
\la{coordchange}
\een
where in this context we use ${}\dot{}{}$ to denote differentiation with respect
to the transformed coordinate $\tlu$: $\dot{u} = \ud u / \ud \wt{u}$.  Under
these coordinate changes, the metric becomes
\bea
\ds & = & \ud \wt{\gr}^2 + 2 \, 
\ud \tlu \,  \ud \tlv + (\tlv^2 \tlf_2 + \tlv \tlf_1 + \tlf_0) \, 
\ud \tlu^2 + 2 ( \tlv \tW_1 + \tW_0 ) \,  \ud \tlu \,  \ud \wt{\gr} \, ,
\eea
with
\bea
\tW_1 & = & W_1 \, , \nnr
\tW_0 & = & \dot{u} (W_0 + W_1 F) + \dot{u}
\fr{\pd F}{\pd \wt{\gr}} + \fr{\ud G}{\ud \tlu} \, , \nnr
\tlf_2 & = & f_2 \, , \nnr
\tlf_1 & = & \dot{u} (f_1 + 2 f_2 F) + 2 W_1 \fr{\ud G}{\ud \tlu} -
\fr{2 \ddot{u}}{\dot{u}} \, , \nnr
\tlf_0 & = & \dot{u}^2 (f_0 + f_1 F + f_2 F^2) +
2 \dot{u} \fr{\pd F}{\pd \tlu} + 2 \dot{u} (W_0 + W_1 F)
\fr{\ud G}{\ud \tlu} + \lt( \fr{\ud G}{\ud \tlu} \rt)^2 \, .
\eea

Next, the $v u$ or $\gr \gr$ component of the field equation gives, 
written in a suggestive way,
\ben
(W_1' - \tf{1}{2} W_1^2 - 2 \gL)' + (\mu - \tf{3}{2} W_1) (W_1' - \tf{1}{2}
W_1^2 - 2 \gL) = 0 \, ,
\la{W1Kundt}
\een
from which we can solve for $W_1 (u, \gr)$.  The general solution involves two constants of integration, but if $W_1 (\gr)$ is a solution, then so too is $W_1 (\gr + c(u))$, for arbitrary $c(u)$.  Therefore one constant of integration can be removed by the coordinate transformation $\gr \ra \gr - c(u)$, provided that $W_1$ is not independent of $\gr$.  Two particularly simple classes of solutions are $W_1 = \tf{2}{3} \mu$ and $W_1' = \tf{1}{2} W_1^2 + 2 \gL$; we shall study these in more detail later.  Having
obtained $W_1$, a coordinate transformation of the form $\tlv = v + F(u, \gr)$ means that we can set $W_0 = 0$.  Then the $u \gr$ component of the field equation gives
\ben
f_1 '' + (\mu - \tf{1}{2} W_1) f_1' = 0 \, , 
\la{f1Kundt}
\een
from which we can obtain $f_1 (u, \gr)$.  The general solution is of the form $f_1(u,
\gr) = F(\gr) f_{11} (u) + f_{12} (u)$, with $F'$ not identically zero, and
where $f_{11}$ and $f_{12}$ are arbitrary.  By a coordinate transformation of
the form $\tlv = v / \dot{u}(\tlu)$, $u = u (\tlu)$, we can set $f_{12} = 0$.  
Then the $u u$ component of the field equation gives
\bea
&& f_0''' + (\mu + \tf{3}{2} W_1) f_0'' + \tf{1}{4} (14 W_1' + 4 \mu W_1 - W_1^2
- 12 \gL) f_0' \nnr
&& + (-\mu W_1' + \tf{9}{2} W_1 W_1' 
- 6 \gL W_1 + \mu W_1^2 - \tf{3}{2} W_1^3 + 4 \gL \mu) f_0 
= - F' (\dot{f}_{11} + \tf{1}{2} F f_{11}^2) \, ,
\la{f0Kundt}
\eea
a linear ordinary differential equation, from which we can obtain $f_0 (u, \gr)$.

In summary, a Kundt spacetime that is a solution of TMG can be put in the form 
(\ref{Kundtmetric}), with
\bea
P (u, \gr) & = & 1 \, , \nnr
W (v, u, \gr) & = & v W_1 (u, \gr) + W_0 (u, \gr) \, , \nnr
f (v, u, \gr) & = & v^2 f_2 (u, \gr) + v f_1 (u, \gr) + f_0 (u, \gr) \, ,
\la{Wf}
\eea
where $W_1$ solves (\ref{W1Kundt}) and, provided that $W_1$ is not independent of $\gr$, can have a constant of integration removed by $\gr \ra \gr - c(u)$, $W_0$ can be chosen to vanish, $f_2$ is given by (\ref{f2Kundt}), $f_1$ solves
(\ref{f1Kundt}) and can be chosen so that $f_1 = F(\gr) f_{11} (u)$ with $F'
\neq 0$, and $f_0$ solves (\ref{f0Kundt}).  There is still some coordinate
freedom remaining, which we shall use in different ways when considering
specific examples later.

Note that, for each allowable choice of $W_1$, there
is always a solution by taking $f_0 = 0$ and $f_1 = 0$.  A more general
solution, with $f_0$ and $f_1$ turned on, can be regarded as being of
generalized Kerr--Schild form, as the $f_0 = 0$, $f_1 = 0$ ``background'' metric
plus the square of a 1-form that is null with respect to both the background
metric and the full metric.


\sse{Special Kundt solutions}


We remarked that two particularly simple classes of solutions of (\ref{W1Kundt}) are $W_1 = \tf{2}{3} \mu$ and $W_1' = \tf{1}{2} W_1^2 + 2 \gL$.  In the latter case, $W_1$ can be chosen to be a function of $\gr$ only.  In Sections 3 and 4 respectively, we give such solutions explicitly, by explicitly solving the differential equations.  Before doing so, however, we elaborate on why these solutions are special.


\ssse{Algebraic classification}


Algebraic classification of curvature is a useful tool for classifying spacetimes independent of coordinate systems.  It has been used to show that many solutions in the TMG literature are locally equivalent by coordinate transformations \cite{chposeI}.  For the formulation of the Petrov--Segre classification that we use here, see \cite{chposeI}, in particular Table 1.  In brief, it can be regarded as algebraic classification of the linear map given by the traceless Ricci tensor $S{^\mu}{_\nu} = R{^\mu}{_\nu} - \tf{1}{3} R \gd^\mu_\nu$, or equivalently of the map $C{^\mu}{_\nu}$, according to Jordan normal form.  It is helpful to find the Petrov--Segre types for the Kundt solutions of TMG that we have just found, to reassure ourselves that they are indeed new, and to single out the most algebraically special solutions, which may be of more utility.

For the general Kundt solution, the eigenvalues of $S{^\mu}{_\nu}$ 
are $\ga$ and $- 2 \ga$ with algebraic multiplicities 2 and 1 
respectively, where
\ben
\ga = - \tf{1}{2} (W_1' - \tf{1}{2} W_1^2 - 2 \gL) \, .
\een
Because of the repeated eigenvalue, no such solution can have Petrov--Segre type I.  The solution being of Petrov--Segre type III or N (or O) is equivalent to $\ga = 0$, i.e.~$W_1' = \tf{1}{2} W_1^2 + 2 \gL$.  The remaining solutions are of type D or II.  Equivalently, we can see these Petrov--Segre types from the scalar invariants $I := S{^\mu}{_\nu} S{^\nu}{_\mu} = 6 \ga^2$, $J := S{^\mu}{_\nu} S{^\nu}{_\gr} S{^\gr}{_\mu} = - 6
\ga^3$; type I is excluded because $I^3 =	 6 J^2$.  To more precisely determine the Petrov--Segre type, we can compute the minimal polynomial of $S{^\mu}{_\nu}$, comparing with the final column of Table 1 of \cite{chposeI}.  In particular,
\ben
M_{\mu \nu} := S{_\mu}{^\gr} S_{\gr \nu} + \ga S_{\mu \nu} - 2 \ga^2 g_{\mu \nu}
\een
vanishes for types D and N, but does not vanish for types II and III.  For the general Kundt solution, only the
$uu$ component does not vanish:
\ben
M_{uu} = \fr{v \ga}{8} (2 \mu - 3 W_1) (\mu v \ga + F' f_{11}) + \fr{3 \ga}{4}
(f_0'' + W_1 f_0' + W_1' f_0) + \fr{1}{4} (F')^2 f_{11}^2 \, .
\la{Kundtpoly}
\een
From the $v$ or $v^2$ coefficient we see that a type D Kundt solution
necessarily has $W_1 = \tf{2}{3} \mu$.  Also, taking $\ga = 0$, i.e.~$W_1' =
\tf{1}{2} W_1^2 + 2 \gL$, we see that type N solutions have $f_{11} = 0$ and
type III solutions have $f_{11} \neq 0$.  In summary, a non-trivial Kundt solution is:
\begin{itemize}
\item Type D only if $W_1 = \tf{2}{3} \mu$ and $\gL \neq - \tf{1}{9} \mu^2$.
\item Type N if and only if $W_1' = \tf{1}{2} W_1^2 + 2 \gL$ and $f_{11} = 0$.
\item Type III if and only if $W_1' = \tf{1}{2} W_1^2 + 2 \gL$ and $f_{11} \neq 0$.
\item Type II otherwise.
\end{itemize}

Note that Kundt solutions of TMG are never type I.  In four-dimensional vacuum Einstein gravity, type I Kundt solutions are ruled out by the Goldberg--Sachs theorem \cite{goldsach}, which is an important result connecting algebraic classification and null geodesic congruences.  This states that a vacuum solution of Einstein gravity (without a cosmological constant) is algebraically special, i.e.~more special than type I, if and only if it admits a null geodesic congruence that is shear-free.  A counter-example to this type of result in TMG is provided by the triaxially squashed AdS$_3$ solutions \cite{nutbak, ortiz}, which are of Petrov--Segre type I$_\bbR$ but whose null geodesic congruences, as for any three-dimensional spacetime, are shear-free.  Perhaps there is a modified theorem that would apply to TMG and rule out type I Kundt solutions.


\ssse{Constant scalar invariant spacetimes}


A scalar polynomial curvature invariant is a scalar that is constructed as a
polynomial in the curvature and its covariant derivatives.  Such scalar
invariants provide coordinate-independent information that may distinguish
different geometries.  A spacetime for which all the scalar polynomial
invariants are constant is known as a constant scalar invariant (CSI) spacetime
\ci{cohepe1}.

A common property of the biaxially squashed AdS$_3$ and AdS pp-wave solutions is that they are CSI spacetimes.  For the biaxially squashed AdS solutions, we
can see this from $S_{\mu \nu} = (\tf{1}{9} \mu^2 - m^2) (g_{\mu \nu} \pm 3 k_\mu k_\nu)$ and $\na_\mu k_\nu = \tf{1}{3} \mu \gep_{\mu \nu \gr} k^\gr$: any scalar polynomial curvature invariant reduces to a constant coefficient polynomial in $k^\mu$, which is a unit-normalized vector, so is constant.  For the AdS pp-wave
solution, we can see this from being able to write $S_{\mu \nu}$ in the form $S_{\mu \nu} = c f_1(u) \ue^{-(3m+\mu)\rho} k_\mu k_\nu$, where $c$ is some constant and $k^\mu$ is a null Killing vector, with $\na_\mu k_\nu = - m \gep_{\mu \nu \gr} k^\gr$ \ci{gipose}.

All three-dimensional CSI spacetimes are known \ci{cohepe2}, and an initial
classification splits these into two classes: locally homogeneous spacetimes, and Kundt spacetimes (or possibly both).  A three-dimensional CSI Kundt metric can be written in the form (\ref{Kundtmetric}), with $W$ and $f$ of the
form (\ref{Wf}), and with
\bea
&& f_2 = \gs + \tf{1}{4} W_1^2 \, , \la{f2CSI} \\
&& W_1' - \tf{1}{2} W_1^2 = s \, , \la{W1CSI} \\
&& (2 \gs - s) W_1 = 2 \ga  \, , \la{alphaCSI}
\eea
for some constants $\gs$, $s$ and $\ga$.  The classification of CSI Kundt
spacetimes then splits into two cases: $s = 2 \gs$ and $s \neq 2 \gs$.

We now demand that these geometries give solutions of TMG.  From
(\ref{f2Kundt}), (\ref{f2CSI}) and (\ref{W1CSI}), we have $s = 3 \gL - \gs$.  If
$s = 2 \gs$, then $s = 2 \gL$, and so (\ref{W1CSI}) gives $W_1'= \tf{1}{2} W_1^2
+ 2 \gL$.  If instead $s \neq 2 \gs$, then (\ref{alphaCSI}) implies that $W_1$
is a constant, which, from (\ref{W1Kundt}), must be $W_1 = \tf{2}{3} \mu$.


\ssse{Summary}


We have seen that the families of solutions with $W_1 = \tf{2}{3} \mu$ and with $W_1' = \tf{1}{2} W_1^2 + 2 \gL$ are privileged.  If $\gL = - \tf{1}{9} \mu^2$, then, in light of our discussion of algebraic types, $W_1 = \tf{2}{3} \mu$ is better thought of as a special case of  $W_1' = \tf{1}{2} W_1^2 + 2 \gL$.  The two families are characterised as the  Kundt solutions that are CSI: the $W_1 = \tf{2}{3} \mu$ family is generically of Petrov--Segre type II, and the $W_1' = \tf{1}{2} W_1^2 + 2 \gL$ family is generically of type III.  In more detail, we have:

\pa{\underline{$W_1 = \tf{2}{3} \mu$ ($\gL \neq - \tf{1}{9} \mu^2$); \bf
Petrov--Segre Type II}:}

These CSI Kundt solutions are generically of Petrov--Segre type II,
but in special cases are type D.  The type D solutions are biaxially spacelike-squashed AdS$_3$, of which the general type II CSI Kundt solutions can be considered deformations.  In a slightly more refined algebraic classification, the type D solutions would be classed as type $D_\us$, since the one-dimensional eigenspace of $S{^\mu}{_\nu}$ is spacelike.

\pa{\underline{$W_1' = \tf{1}{2} W_1^2 + 2 \gL$; \bf Petrov--Segre Type III}:}

These CSI Kundt solutions are generically of Petrov--Segre type III, 
but in special cases are type N (or O).  All type N and type III Kundt 
solutions are CSI.  There are 6 cases for $W_1$ to consider, with various signs
of the cosmological constant.  Unlike in Type II, here it is necessary to
distinguish between the three choices of negative, zero or positive cosmological
constant, since we shall find families of solutions that merge or disappear with
these choices.  These solutions can be considered deformations of the round AdS$_3$,
which is type O.

\begin{enumerate}

\item {\bf Negative cosmological constant:} There are 3 
cases: $W_1 (\gr) = - 2 m$, $W_1 (\gr) = 
 - 2 m \coth (m \gr)$, $W_1 (\gr) = - 2 m \tanh (m \gr)$.  
Note that if $\gr \ra \gr + \ui \pi / 2 m$, then $\sinh (m \gr) \leri
\cosh (m \gr)$, so the second and third cases here are related by
analytic continuation.

\item {\bf Zero cosmological constant:} There are 2 cases: $W_1 (\gr) = 0$, $W_1 (\gr) = - 2 / \gr$.  If we take the zero  cosmological constant limit $m \ra 0$ of the negative cosmological constant cases, then $W_1 = - 2 m$ and $W_1 = - 2 m \tanh (m \gr)$ both lead to $W_1 = 0$, whereas $W_1 = - 2 m \cot (m \gr)$ leads to $W_1 = - 2 / \gr$.

\item {\bf Positive cosmological constant:} Suppose that we instead 
consider a positive cosmological constant, by
replacing $m \ra \ui m$.  Since $W_1 = - 2 m$ does not lead to real
$W_1$, and since also $\tan (\fr{\pi}{2} - x) = \cot x$, there is
essentially 1 case: $W_1 (\gr) = - 2 m \cot (m \gr)$.

\end{enumerate}

To visualise why we have various branches of solutions in the $W_1' = \tf{1}{2}
W_1^2 + 2 \gL$ family, we can consider a phase portrait for the $W_1$
differential equation (\ref{W1Kundt}).  By defining $x = W_1 / \mu$ and $y =
W_1' / \mu$, we have
\ben
x' = y \, , \qd y' = \mu x y - (1 - \tf{3}{2} x) (\mu y - \tf{1}{2} \mu^2 x^2 - 2
\gL) \, .
\een
There is a fixed point at $(x, y) = (\tf{2}{3}, 0)$, corresponding to 
the $W_1 = \tf{2}{3} \mu$ solution.  Another solution is $y = \tf{1}{2} \mu x^2
+ 2 \gL / \mu$, corresponding to the $W_1' = \tf{1}{2} W_1^2 + 2 \gL$ family. 
This describes a parabola in the $x y$-plane.  For $\gL < 0$, the parabola
intersects the $x$-axis at $(x, y) = (\pm 2 \sr{- \gL} / \mu, 0)$, where the
flow has a fixed point and corresponds to the solution $W_1 = - 2 m$ (the choice
of signs corresponds to allowing $m$ to have any sign).  The part of the
parabola in between these fixed points corresponds to $W_1 = - 2 m \tanh (m
\gr)$, and the remainder of the parabola corresponds to $W_1 = - 2 m \coth (m
\gr)$.  For $\gL > 0$, there is no intersection with the $x$-axis, and so only
one branch of solutions.

Note the analogy with four-dimensional Einstein gravity.  In four
dimensions, the type N solutions with a cosmological constant have been obtained
in \cite{garple, garcia, ozroro} (see \cite{bicpod} for a summary).  The
families of solutions have a similar structure: there are three families for
$\gL < 0$, two families for  $\gL = 0$, and one family for $\gL > 0$.


\se{CSI Type II: Deformations of Spacelike-Squashed AdS$_3$}


Here, we shall explicitly obtain, by solving the differential equations that
arise from the field equation, solutions of TMG that have Petrov--Segre type II
CSI Kundt metrics.  Any type D Kundt solution must be of this form, and we find
that it must be spacelike-squashed AdS$_3$.  The general type II solutions here
are deformations of spacelike-squashed AdS$_3$.


\sse{Solutions}


From (\ref{f1Kundt}), we have
\ben
f_1'' + \tf{2}{3} \mu f_1' = 0 \, ,
\een
and so
\ben
f_1 (u, \gr) = \ue^{- 2 \mu \gr / 3} f_{11} (u) \, ,
\een
taking $f_{12} = 0$, as in the general case.  The coordinate
transformation $v = \tlv - \tf{1}{2} (\tf{1}{3} \mu^2 + 3 \gL)^{-1}
\ue^{- 2 \mu \gr / 3} f_{11} (u)$ allows us to set $f_{11} = 0$,
and so $f_1 = 0$.  From (\ref{f0Kundt}), we have
\ben
f_0''' + 2 \mu f_0'' + (\tf{5}{9} \mu^2 - 3 \gL) f_0' = 0 \, .
\la{f0typeII}
\een
The general solution of TMG that follows is
\ben
\ds = 2 \,  \ud u \,  \ud v - \tf{1}{9} (\mu^2 - 27 \gL) v^2 \, 
\ud u^2 + (\ud \gr + \tf{2}{3} \mu v \,  \ud u)^2 + f_0 (u, \gr) \,  \ud u^2 \, ,
\label{masterII}
\een
where $f_0$ satisfies (\ref{f0typeII}).


\pa{Negative cosmological constant:}


For a negative cosmological constant $\gL = - m^2$, we have
\begin{align}
& \underline{\mu^2 > \tf{27}{4} m^2}: & f_0 (u, \gr) & =
\ue^{- \mu \gr} \cosh (\gc m \gr) f_{01} (u) + \ue^{- \mu \gr}
\sinh (\gc m \gr) f_{02} (u) + f_{03} (u) \, , \nnr
&& \gc & = \sr{\fr{4 \mu^2}{9 m^2} - 3} \, , \la{defsquads} \\
& \underline{\mu^2 < \tf{27}{4} m^2}: & f_0 (u, \gr) & =
\ue^{- \mu \gr} \cos (\gc m \gr) f_{01} (u) +
\ue^{- \mu \gr} \sin (\gc m \gr) f_{02} (u) + f_{03} (u) \, , \nnr
&& \gc & = \sr{3 - \fr{4 \mu^2}{9 m^2}} \, , \\
& \underline{\mu^2 = \tf{27}{4} m^2}: & f_0 (u, \gr) & =
\gr \ue^{- \mu \gr} f_{01} (u) + \ue^{- \mu \gr} f_{02} (u) + f_{03} (u) \, ,
\end{align}
with $f_{01}$, $f_{02}$ and $f_{03}$ arbitrary.  
However, we can set $f_{03} = 0$ by a coordinate transformation
of the form (\ref{coordchange}).  Specifically, we take
\ben
F(\tlu, \wt{\gr}) = - \fr{3}{2 \mu \dot{u}} \fr{\ud G}{\ud \tlu} = - \fr{9
\ddot{u}}{(\mu^2 + 27 m^2) \dot{u}^2} \, ,
\een
which preserves $W_0 = 0$ and $f_1 = 0$, and take $u(\tlu)$ to 
satisfy the differential equation that removes $f_{03}$:
\ben
\{ u, \tlu \} = - \tf{1}{18} (\mu^2 + 27 m^2) \dot{u} f_{03} (u) \, ,
\een
where
\ben
\{ u, \tlu \} := \fr{\dddot{u}}{\dot{u}} - \fr{3 \ddot{u}^2}{2 \dot{u}^2}
\la{Schwarzian}
\een
is the Schwarzian derivative of $u$ with respect to $\tlu$.


\pa{Zero cosmological constant:}


Taking $m = 0$ in (\ref{defsquads}), we obtain the zero cosmological
constant solution, which can be given in the form (\ref{masterII}), where
\ben
f_0(u,\rho) =
\ue^{- \mu \gr / 3} f_{01} (u) + \ue^{- 5 \mu \gr / 3} f_{02} (u) +
f_{03} (u)  \, ,
\een
with $f_{01}$, $f_{02}$ and $f_{03}$ arbitrary.  Again, we can set 
$f_{03} = 0$ by coordinate transformations.


\pa{Positive cosmological constant:}


For a positive cosmological constant $\gL = m^2$, we take $m \ra \ui m$ in
(\ref{defsquads}), giving the solution (\ref{masterII}), where
\bea
f_0(u,\rho)& = &\ue^{- \mu \gr} \cosh (\gc m \gr) f_{01} (u) +
\ue^{- \mu \gr} \sinh (\gc m \gr) f_{02} (u) + f_{03} (u) 
\, , \nnr
\gc & = & \sr{\fr{4 \mu^2}{9 m^2} + 3} \, .
\eea
with $f_{01}$, $f_{02}$ and $f_{03}$ arbitrary.  
Again, we can set $f_{03} = 0$ by coordinate transformations.


\sse{Properties}


The solutions are generically of Petrov--Segre type II, but in special cases are
type D.  If a solution is type D, then (\ref{Kundtpoly}) vanishes, and so $f_0''
+ \tf{2}{3} \mu f_0' = 0$.  Using (\ref{f0typeII}) and $\gL \neq - \tf{1}{9}
\mu^2$, we see that a type D Kundt solution has $f_0' = 0$, and so, for the
solutions explicitly given above, $f_0 = 0$.

We may also see the Petrov--Segre type by using the canonical forms of the
traceless Ricci tensor, which are given in the third column of Table 1 of \cite{chposeI}.  In this case, we have
\ben
S_{\mu \nu} = (\tf{1}{9} \mu^2 + \gL) (g_{\mu \nu} - 3 s_\mu s_\nu)
+ q k_\mu k_\nu \, ,
\la{defsAdSS}
\een
where $s$ is the unit spacelike vector $s^\mu \pd_\mu = \pd / \pd \gr$, which
has associated 1-form $s_\mu \,  \ud x^\mu = \ud \gr + \tf{2}{3} \mu v \,  \ud u$,
and $k$ is the null vector $k^\mu \pd_\mu = \pd / \pd v$, which has associated
1-form $k_\mu \,  \ud x^\mu = \ud u$, and which is orthogonal to $s$.  The scalar
function $q$ is
\ben
q = \tf{2}{3} \mu f_0' + (\tf{5}{18} \mu^2 - \tf{3}{2} \gL) f_0 \, ,
\een
where $f_0$ has $f_{03} = 0$.  From the form of $S_{\mu \nu}$, we see that the
solution generically has Petrov--Segre type II.  In the special case of $f_0 =
0$, the solution has Petrov--Segre type D$_\us$.

The covariant derivatives of $s$ and $k$ are
\ben
\na_\mu s_\nu = \tf{1}{3} \mu \gep_{\mu \nu \gr} s^\gr +
\tf{1}{2} f_0' k_\mu k_\nu \, ,
\een
\ben
\na_\mu k_\nu = \tf{1}{3} \mu (k_\mu s_\nu + s_\mu k_\nu)
- \tf{1}{9} (\mu^2 - 27 \gL) v k_\mu k_\nu \, .
\een
Using (\ref{defsAdSS}), these derivatives, and that $f_0'' + 2 \mu
f_0' + (\tf{5}{9} \mu^2 - 3 \gL) f_0 = 0$, we can check that the field
equation (\ref{fieldeq2}) is solved.

The type D solution, or equivalently the $f_0 = 0$ solution, describes spacelike-squashed AdS$_3$.  We can explicitly see this by transforming the metric, which in this case is
\ben
\ds = 2 \,  \ud u \,  \ud v - \tf{1}{9} (\mu^2 + 27 m^2) v^2 \, 
\ud u^2 + (\ud \gr + \tf{2}{3} \mu v \,  \ud u)^2 \, .
\la{typeIIsAdS}
\een
Making the coordinate changes $\hat{u} = \tf{1}{18} (\mu^2 + 27 m^2)
u$, $\hat{v} = 1/v + \tf{1}{18} (\mu^2 + 27 m^2) u$, we have
\ben
\ds = - \fr{36}{(\mu^2 + 27 m^2) (\hat{u} - \hat{v})^2} \, 
\ud \hat{u} \,  \ud \hat{v} + \lt( \ud \gr - \fr{12 \mu}{\mu^2 + 27 m^2}
\fr{\ud \hat{u}}{\hat{u} - \hat{v}} \rt) ^2 \, .
\een
Then making the coordinate changes $t = \half (\hat{u} + \hat{v})$, $x
= \half (\hat{v} - \hat{u})$, $z = \tf{1}{6} \mu^{-1} (\mu^2 + 27 m^2)
\gr - \log (\hat{v} - \hat{u})$, the solution is
\ben
\ds = \fr{9}{\mu^2 + 27 m^2} \bigg[ \fr{- \ud t^2 +
\ud x^2}{x^2} + \fr{4 \mu^2}{\mu^2 + 27 m^2} \lt( \ud z +
\fr{\ud t}{x} \rt) ^2 \bigg] \, ,
\een
which is manifestly spacelike-squashed AdS$_3$, written using Poincar\'{e} coordinates $(t, x)$ for the AdS$_2$ part.  Other coordinate systems used in the literature can be found in \cite{chposeI}.

Therefore the general Petrov--Segre type II CSI Kundt solution 
can be regarded as a deformation of 
spacelike-squashed AdS$_3$.  Furthermore, it can be regarded as of 
generalized Kerr--Schild form, with a spacelike-squashed AdS$_3$ background
metric.


\se{CSI Type III: Deformations of Round AdS$_3$}


Here, we shall explicitly obtain solutions of TMG that have Petrov--Segre type 
III CSI Kundt metrics.  Any type N or type III Kundt solution must be of this
form.  The general solutions here are deformations of the round AdS$_3$.  For a
negative cosmological constant, there are three families of solutions: $W_1 = -
2 m$, $W_1 = - 2 m \coth (m \gr)$, and $W_1 = - 2 m \tanh (m \gr)$.  The $W_1 =
- 2 m$ family includes the AdS pp-wave solution.  The other two families are
similar, and so we shall be more brief in their discussion.  Where applicable,
we shall also consider the extension of these families to zero or positive
cosmological constant.


\sse{The $W_1 = - 2 m$ family}



\ssse{Solutions}



\pa{Negative cosmological constant:}


From (\ref{f1Kundt}), we have
\ben
f_1'' + (\mu + m) f_1' = 0 \, .
\een
We consider separately the cases $\mu \neq \pm m$, $\mu = m$ and $\mu = - m$.


\pa{\underline{$\mu \neq \pm m$}:}


If $\mu \neq \pm m$, then
\ben
f_1 (u, \gr) = \ue^{- (m + \mu) \gr} f_{11} (u) \, ,
\een
taking $f_{12} = 0$, as in the general case.  From (\ref{f0Kundt}), we have
\ben
f_0''' + (\mu - 3 m) f_0'' + 2 m (m - \mu)
f_0' = (m + \mu) (\ue^{- (m + \mu) \gr} \dot{f}_{11} +
\tf{1}{2} \ue^{- 2 (m + \mu) \gr} f_{11}^2) \, .
\een
The general solution is
\ben
f_0 (u, \gr) = \ue^{(m - \mu) \gr} f_{01} (u) + \ue^{2 m \gr} f_{02} (u) +
f_{03} (u) - \fr{\ue^{- (m + \mu) \gr} \dot{f}_{11} (u)}{2 m (\mu + 3m)}
- \fr{\ue^{- 2 (m + \mu) \gr} f_{11}(u)^2}{8 (\mu + 2m) (\mu + 3m)} \, .
\een
By a coordinate transformation of the form $v = \tlv + F(u) \ue^{2 m
  \gr}$, we can set $f_{02} = 0$.  We can also set $f_{03} = 0$ by a coordinate 
transformation of the form (\ref{coordchange}).  Specifically, we have
\ben
F(\tlu, \wt{\gr}) = \fr{1}{2 m \dot{u}} \fr{\ud G}{\ud \tlu} = 
- \fr{\ddot{u}}{4 m^2 \dot{u}^2} \, ,
\een
which preserves $W_0 = 0$ and $f_{12} = 0$, and take $u(\tlu)$ to 
satisfy the differential equation that removes $f_{03}$:
\ben
\{ u, \tlu \} = 2 m^2 \dot{u} f_{03} (u) \, ,
\een
where $\{ u, \tlu \}$ is the Schwarzian derivative (\ref{Schwarzian}).  
The general solution of TMG that follows is
\ben
\ds = \ud \gr^2 + 2 \,  \ud u \,  \ud v -
4 m v \,  \ud u \,  \ud \gr + [v \ue^{- (m + \mu) \gr} f_{11} (u) +
f_0 (u, \gr)] \,  \ud u^2 \, ,
\een
where
\ben
f_0 (u, \gr) = \ue^{(m - \mu) \gr} f_{01} (u) - \fr{\ue^{- (m + \mu) \gr}
\dot{f}_{11} (u)}{2 m (\mu + 3m)} -
\fr{\ue^{- 2 (m + \mu) \gr} f_{11}(u)^2}{8 (\mu + 2m) (\mu + 3m)} \, .
\een


\pa{\underline{$\mu = m$}:}


The $\mu = m$ solution is
\ben
\ds = \ud \gr^2 + 2 \,  \ud u \,  \ud v -
4 m v \,  \ud u \,  \ud \gr + [v \ue^{- 2 m \gr} f_{11} (u) +
f_0 (u, \gr)] \,  \ud u^2 \, ,
\een
where
\ben
f_0 (u, \gr) = \gr f_{01} (u) - \fr{\ue^{- 2 m \gr}
\dot{f}_{11} (u)}{8 m^2} - \fr{\ue^{- 4 m \gr} f_{11}(u)^2}{96 m^2} \, .
\een


\pa{\underline{$\mu = -m$}:}


If $\mu = -m$, then
\ben
f_1 (u, \gr) = \gr f_{11} (u) \, .
\een
From (\ref{f0Kundt}), we have
\ben
f_0''' - 4 m f_0'' + 4 m^2 f_0' = - (\dot{f}_{11} + \tf{1}{2} \gr f_{11}^2) \, .
\een
After again removing two redundant functions of $u$ by coordinate
transformations, we obtain the solution
\ben
\ds = \ud \gr^2 + 2 \,  \ud u \,  \ud v - 4 m v \, 
\ud u \,  \ud \gr + [v \gr f_{11} (u) + f_0 (u, \gr)] \,  \ud u^2 \, ,
\een
where
\ben
f_0 (u, \gr) = \gr \ue^{2 m \gr} f_{01} (u) -
\fr{\gr \dot{f}_{11}(u)}{4 m^2} - \fr{\gr (m \gr + 2) f_{11}(u)^2}{16 m^3} \, .
\een


\pa{Zero cosmological constant:}


The zero cosmological constant solution is
\ben
\ds = \ud \gr^2 + 2 \,  \ud u \,  \ud v + [v \ue^{-\mu \gr} f_{11}(u) +
\ue^{- \mu \gr} f_{01} (u) + \mu^{-1} \dot{f}_{11}(u) \gr \ue^{- \mu \gr} -
\tf{1}{8} \mu^{-2} f_{11}(u)^2 \ue^{- 2 \mu \gr}] \,  \ud u^2 \, .
\een


\pa{Positive cosmological constant:}


There is no positive cosmological constant analogue, since sending $m \ra \im m$
would give complex $W_1$.


\ssse{Properties}


The solutions are generically of Petrov--Segre type III, but in special cases
are type N (or O).  If a solution is type N, then (\ref{Kundtpoly}) vanishes,
with $\ga = 0$, and so $f_1' = 0$, and so $f_{11} = 0$.

We may also see the Petrov--Segre type by decomposing the traceless Ricci
tensor.  The traceless Ricci tensor is of the form
\ben
S_{\mu \nu} = k_\mu m_\nu + m_\mu k_\nu \, .
\la{defAdSS}
\een
$k$ is the null vector $k^\mu \pd_\mu = \pd / \pd v$, which has
associated 1-form $k_\mu \,  \ud x^\mu = \ud u$.  We define the unit
spacelike vector $s^\mu \pd_\mu = \pd / \pd \gr$, which has associated
1-form $s_\mu \,  \ud x^\mu = \ud \gr - 2 m v \,  \ud u$.  The spacelike
vector $m$ is
\ben
m_\mu = \tf{1}{2} (m + \mu) (- \ue^{-(m + \mu) \gr} f_{11} s_\mu + q k_\mu) \, ,
\een
where
\ben
q = \fr{1}{2} \bigg( (m - \mu) \ue^{(m - \mu) \gr}
f_{01} + \fr{\ue^{-(m + \mu) \gr} \dot{f}_{11}}{2 m} +
\fr{\ue^{- 2 (m + \mu) \gr} f_{11}^2}{2 (\mu + 3m)} -
(\mu + 3m) v \ue^{- (m + \mu) \gr} f_{11} \bigg) \, .
\een
If $f_{11} \neq 0$, then the solution has Petrov--Segre type III.  If $f_{11} =
0$ and $f_{01} \neq 0$, then the solution has Petrov--Segre type N.  If $f_{11}
= 0$ and $f_{01} = 0$, then the solution has Petrov--Segre type O.

The covariant derivatives of $s$ and $k$ are
\ben
\na_\mu s_\nu = - m \gep_{\mu \nu \gr} s^\gr + \tf{1}{2} f_0' k_\mu k_\nu \, ,
\een
\ben
\na_\mu k_\nu = - m (k_\mu s_\nu + s_\mu k_\nu) +
(\tf{1}{2} \ue^{-(m + \mu) \gr} f_{11} - 4 m^2 v) k_\mu k_\nu \, .
\een
Using (\ref{defAdSS}), these derivatives, and the differential 
equations for $f_0$ and $f_1$, we can check that the field
equation (\ref{fieldeq2}) is solved.

The $f_{11} = 0$, $f_0 = 0$ solution is the round AdS$_3$ (or flat spacetime for $m
= 0$).  The metric is
\ben
\ds = \ud \gr^2 + 2 \,  \ud u \,  \ud v - 4 m v \,  \ud u \,  \ud \gr \, ;
\een
making the coordinate change $v \ra v \ue^{2 m \gr}$, we have
\ben
\ds = \ud \gr^2 + 2 \ue^{2 m \gr} \,  \ud u \,  \ud v \, ,
\een
which is recognisable as AdS$_3$.

Therefore the general solution can be regarded as a deformation of the round
AdS$_3$.  
Furthermore, it can be regarded as of 
generalized Kerr--Schild form, with a round AdS$_3$ background metric.


\sse{The $W_1 = - 2 m \coth (m \gr)$ family}



\ssse{Solutions}



\pa{Negative cosmological constant:}


From (\ref{f1Kundt}), we have
\ben
[\ue^{\mu \gr} \sinh (m \gr) f_1']' = 0 \, ,
\een
and so\footnote{Here and elsewhere, if there are divergent integrals, then a
finite result can be obtained by reinstating redundant functions, in this case
$f_{12}$.}
\ben
f_1 (u, \gr) = F(\gr) f_{11} (u) \, , \qd F(\gr) = m \int^\gr_\infty \ud t \, 
\fr{\ue^{- \mu t}}{\sinh (m t)} \, .
\een
We consider separately the
cases $\mu \neq \pm m$, $\mu = m$, $\mu = - m$.

\pa{\underline{$\mu \neq \pm m$}:}

From (\ref{f0Kundt}), we have
\bea
&& \lt( \fr{f_0}{1 - \ue^{- 2 m \gr}} \rt) ''' + (\mu - 3m)
\lt( \fr{f_0}{1 - \ue^{- 2 m \gr}} \rt) '' +
2 m (m - \mu) \lt( \fr{f_0}{1 - \ue^{- 2 m \gr}} \rt) ' \nnr
&& = - \fr{m \ue^{(m - \mu) \gr}}{2 \sinh^2 (m \gr)} (\dot{f}_{11} + \tf{1}{2} F
f_{11}^2) \, .
\eea
The general solution of TMG that follows is
\bea
\ds & = & \ud \gr^2 + 2 \,  \ud u \,  \ud v + m^2 v^2 \cosech^2 (m \gr)
\,  \ud u^2 - 4 m v \coth (m \gr) \,  \ud u \,  \ud \gr \nnr
&& + [v F f_{11} (u) + f_0 (u, \gr)] \,  \ud u^2 \, ,
\eea
where
\bea
\fr{f_0 (u, \gr)}{1 - \ue^{-2 m \gr}} & = & \ue^{(m - \mu) \gr} f_{01} (u) +
\fr{\dot{f}_{11}(u)}{2 m} \int_\infty^\gr \ud t \,  \ue^{2 m t} F(t) \nnr
&& - \fr{m f_{11}(u)^2}{4} \int_\infty^\gr \ud t \,  \int_\infty^t \ud t' \, 
\int_\infty^{t'} \ud t'' \,  \fr{\ue^{2 m (t - t')} \ue^{(m - \mu) t'}
F(t'')}{\sinh^2 (m t'')} \, .
\eea
(We have used coordinate transformations to discard two arbitrary
functions of $u$, corresponding to $f_0(u,\rho) / (1 - \ue^{- 2 m \gr}) =
\ue^{2 m \gr} f_{02} (u) + f_{03} (u)$.)  
We have presented the $f_{11}^2$ coefficient, which does not seem to be
expressible in closed form, 
in integral form by composing inverses of differential operators.  It is
possible to instead give 
the coefficient in terms of a single integral by using a Green function.  


\pa{\underline{$\mu = m$}:}


If $\mu = m$, then
\ben
f_1 (u, \gr) = \log ( 1 - \ue^{- 2 m \gr} ) f_{11} (u) \, .
\een
From (\ref{f0Kundt}), we have
\ben
\lt( \fr{f_0}{1- \ue^{- 2 m \gr}} \rt) ''' - 2 m
\lt( \fr{f_0}{1- \ue^{- 2 m \gr}} \rt) '' = -
\fr{m}{2 \sinh^2 (m \gr)} [\dot{f}_{11} + \tf{1}{2}
\log(1 - \ue^{- 2 m \gr}) f_{11}^2] \, .
\een
We therefore have
\bea
\ds & = & \ud \gr^2 + 2 \,  \ud u \,  \ud v + m^2 v^2
\cosech^2 (m \gr) \,  \ud u^2 - 4 m v \coth (m \gr) \, 
\ud u \,  \ud \gr \nnr
&& + [v \log(1 - \ue^{- 2 m \gr}) f_{11} (u) + f_0 (u, \gr)] \,  \ud u^2 \, ,
\eea
where
\bea
\fr{f_0 (u, \gr)}{1 - \ue^{- 2 m \gr}} & = & \gr f_{01} (u) +
\fr{\dot{f}_{11}(u)}{4 m^2} (\ue^{2 m \gr} - 1) \log(1 - \ue^{- 2 m \gr}) \nnr
&& + \fr{f_{11}(u)^2}{16 m^2} \{ 4 m^2 \gr^2 + (\ue^{2 m \gr} - 1)
[(1 + \log(1 - \ue^{- 2 m \gr}))^2 - 1] + 2 \uL\ui_2 (\ue^{2 m \gr}) \} \, , \nnr
\eea
and where $\uL\ui_2 (x) := \int_x^0 \ud t \,  t^{-1} \log(1 - t)$ is the
dilogarithm.


\pa{\underline{$\mu = - m$}:}


If $\mu = - m$, then
\ben
f_1 (u, \gr) = \log ( \ue^{2 m \gr} - 1 ) f_{11} (u) \, .
\een
From (\ref{f0Kundt}), we have
\bea
&& \lt( \fr{f_0}{1 - \ue^{- 2 m \gr}} \rt) ''' - 4 m
\lt( \fr{f_0}{1- \ue^{- 2 m \gr}} \rt) '' + 4 m^2
\lt( \fr{f_0}{1- \ue^{- 2 m \gr}} \rt) ' \nnr
&& = - \fr{m \ue^{2 m \gr}}{2 \sinh^2 (m \gr)} [\dot{f}_{11} +
\tf{1}{2} \log(\ue^{2 m \gr} - 1) f_{11}^2] \, .
\eea
We therefore have
\bea
\ds & = & \ud \gr^2 + 2 \,  \ud u \,  \ud v + m^2 v^2 \cosech^2 (m \gr)
\,  \ud u^2 - 4 m v \coth (m \gr) \,  \ud u \,  \ud \gr \nnr
&& + [v \log(\ue^{2 m \gr} - 1) f_{11} (u) + f_0 (u, \gr)] \,  \ud u^2 \, ,
\eea
where
\bea
\fr{f_0 (u, \gr)}{1 - \ue^{- 2 m \gr}} & = & \gr \ue^{2 m \gr} f_{01} (u) +
\fr{\dot{f}_{11}(u)}{4 m^2} (\ue^{2 m \gr} - 1) \log(\ue^{2 m \gr} - 1) \nnr
&& + \fr{f_{11}(u)^2}{16 m^2} \{ (\ue^{2 m \gr} - 1)
[(1 + \log(\ue^{2 m \gr} - 1))^2 - 1] + 2
\ue^{2 m \gr} \uL\ui_2 (\ue^{2 m \gr})\} \, . \nnr
&&
\eea


\pa{Zero cosmological constant:}


From (\ref{f1Kundt}), we have
\ben
(\gr f_1')' + \mu \gr f_1' = 0 \, ,
\een
and so
\ben
f_1 (u, \gr) = \uE\ui (- \mu \gr) f_{11} (u) \, ,
\een
where $\uE\ui (x) := - \int^\infty_{-x} \ud t \,  t^{-1} \ue^{- t}$ is
the exponential integral.  From (\ref{f0Kundt}), we have
\ben
\lt( \fr{f_0}{\gr} \rt) ''' + \mu \lt( \fr{f_0}{\gr} \rt) '' =
- \fr{\ue^{- \mu \gr}}{\gr^2} [\dot{f}_{11} + \tf{1}{2} \uE\ui (- \mu \gr)
f_{11}^2] \, .
\een
We therefore have
\ben
\ds = \ud \gr^2 + 2 \,  \ud u \,  \ud v - \fr{4 v}{\gr} \,  \ud u \, 
\ud \gr + \fr{v^2}{\gr^2} \,  \ud u^2 + [v \uE\ui (- \mu \gr) f_{11} (u) +
f_0 (u, \gr) ] \,  \ud u^2 \, ,
\een
with, after using coordinate transformations to discard two arbitrary
functions of $u$,\footnote{This solution bears some
resemblance to equations (60) and (61) of \ci{maccam}.}
\bea
\fr{f_0 (u, \gr)}{\gr} & = & \ue^{- \mu \gr} f_{01} (u) +
\dot{f}_{11}(u) \gr \uE \ui (-\mu \gr) \nnr
&& + \fr{f_{11}(u)^2}{4 \mu} \{ 2 \ue^{-2 \mu \gr} - 4 (1 - \mu \gr) \uE \ui (-
2 \mu \gr) + 4 \ue^{- \mu \gr} \uE\ui(-\mu \gr) + \mu \gr [\uE\ui(-\mu \gr)]^2
\} \, . \nnr
&& 
\eea


\pa{Positive cosmological constant}


The positive cosmological constant analogue is to take $m \ra \im m$,
i.e.~$W_1 = - 2 m \cot (m \gr)$.  From (\ref{f1Kundt}), we have
\ben
[\ue^{\mu \gr} \sin (m \gr) f_1']' = 0 \, ,
\een
and so
\ben
f_1 (u) = F(\gr) f_{11} (u) \, , \qd F(\gr) = m \int^\gr_\infty \ud t \, 
\fr{\ue^{- \mu t}}{\sin(m t)} \, .
\een
From (\ref{f0Kundt}), we have
\bea
&& \lt( \fr{f_0}{\sin(m \gr)} \rt) ''' +
\mu \lt( \fr{f_0}{\sin(m \gr)} \rt) '' +
m^2 \lt( \fr{f_0}{\sin(m \gr)} \rt) ' + \mu
m^2 \fr{f_0}{\sin(m \gr)} \nnr
&& = - \fr{m \ue^{- \mu \gr}}{\sin^2 (m \gr)} (\dot{f}_{11} + \tf{1}{2} F
f_{11}^2) \, .
\eea

The general solution of TMG that follows is
\bea
\ds & = & \ud \gr^2 + 2 \,  \ud u \,  \ud v + m^2 v^2 \cosec^2 (m \gr) \, 
\ud u^2 - 4 m v \cot (m \gr) \,  \ud u \,  \ud \gr \nnr
&& + [v F(\gr) f_{11} (u) + f_0 (u, \gr)]
\,  \ud u^2 \, ,
\eea
where
\bea
\fr{f_0 (u, \gr)}{\sin (m \gr)} & = & \ue^{- \mu \gr} f_{01} (u) + \fr{\sin (m
\gr) F(\gr)}{m^2} \dot{f}_{11}(u) \nnr
&& - \fr{m f_{11} (u)^2}{2} \int^\gr_\infty \ud t \int^t_\infty \ud t'
\int^{t'}_\infty \ud t'' \fr{\ue^{- \mu (\gr - t + t'')} \cos[m (t - 2 t' +
t'')] F(t'')}{\sin^2 (m t'')} \, . \nnr
&&
\eea


\ssse{Properties}


Again, the solution has Petrov--Segre type III for the generic case 
$f_{11} \neq 0$, type N for $f_{11} = 0$ and $f_{01} \neq 0$, and type O for
$f_{11} = 0$ and $f_{01} = 0$, which is maximally symmetric.  The type N
solution 
is distinct from the AdS pp-wave.


\pa{Negative cosmological constant:}


For the $f_{11} = 0$, $f_0 = 0$ solution, making the coordinate change
\ben
\hat{v} = u - \fr{2 \sinh^2 (m \gr)}{m^2 v} \, ,
\een
we have
\ben
\ds = \ud \gr^2 + \fr{4 \sinh^2 (m \gr)}{m^2 (u - \hat{v})^2} \, 
\ud u \,  \ud \hat{v} \, ,
\een
which can be recognised as the round AdS$_3$.  The general solution can again be
regarded as a deformation of the round AdS$_3$.  It takes a generalized Kerr--Schild
form with an AdS$_3$ background metric.


\pa{Zero cosmological constant:}


For the $f_{11} = 0$, $f_0 = 0$ solution, similarly making the coordinate change
\ben
\hat{v} = u - \fr{2 \gr^2}{v} \, ,
\een
we have
\ben
\ds = \ud \gr^2 + \fr{4 \gr^2}{(u - \hat{v})^2} \,  \ud u \,  \ud \hat{v} \, ,
\een
which can be recognised as flat spacetime.


\pa{Positive cosmological constant:}


For the $f_{11} = 0$, $f_0 = 0$ solution, making the coordinate change
\be
\hat{v} = u - \fr{2 \sin^2 (m \gr)}{m^2 v} \, ,
\ee
we have
\be
\ds = \ud \gr^2 + \fr{4 \sin^2 (m \gr)}{m^2 (u - \hat{v})^2} \, 
\ud u \,  \ud \hat{v} \, ,
\ee
which can be recognised as the round dS$_3$.


\sse{The $W_1 = - 2 m \tanh (m \gr)$ family}



\ssse{Solutions}



\pa{Negative cosmological constant:}


The case $W_1 (u, \gr) = - 2 m \tanh (m \gr)$ can be considered by
taking the case $W_1 (u, \gr) = - 2 m \coth (m \gr)$ and making the
analytic continuation $\gr \ra \gr + \ui \pi / 2 m$.  We have
analogously
\ben
f_1 (u , \gr) = F(\gr) f_{11} (u) \, , \qd F(\gr) = m \int^\gr_\infty \ud t \, 
\fr{\ue^{- \mu t}}{\cosh(m t)} \, . 
\een


\pa{\underline{$\mu \neq \pm m$}:}


The $\mu \neq m$ solution is
\bea
\ds & = & \ud \gr^2 + 2 \,  \ud u \,  \ud v - m^2 v^2 \sech^2 (m \gr) \, 
\ud u^2 - 4 m v \tanh (m \gr) \,  \ud u \,  \ud \gr \nnr
&& + [v F(\gr) f_{11} (u) + f_0 (u, \gr)] \,  \ud u^2 \, ,
\eea
where
\bea
\fr{f_0 (u, \gr)}{1 + \ue^{- 2 m \gr}} & = & \ue^{(m - \mu) \gr} f_{01} (u) -
\fr{\dot{f}_{11}(u)}{2 m} \int_\infty^\gr \ud t \,  \ue^{2 m t} F(t) \nnr
&& - \fr{f_{11}(u)^2}{4 m} \int_\infty^\gr \ud t \,  \int_\infty^t \ud t' \, 
\int_\infty^{t'} \ud t'' \,  \fr{\ue^{2 m (t - t')} \ue^{(m - \mu) t'}
F(t'')}{\cosh^2 (m t'')} \, .
\eea


\pa{\underline{$\mu = m$}:}


The $\mu = m$ solution is
\bea
\ds & = & \ud \gr^2 + 2 \,  \ud u \,  \ud v - m^2 v^2 \sech^2 (m \gr) \, 
\ud u^2 - 4 m v \tanh (m \gr) \,  \ud u \,  \ud \gr \nnr
&& + [v \log(1 + \ue^{- 2 m \gr}) f_{11} (u) + f_0 (u, \gr)] \,  \ud u^2 \, ,
\eea
where
\bea
\fr{f_0 (u, \gr)}{1 + \ue^{- 2 m \gr}} & = & \gr f_{01} (u) -
\fr{\dot{f}_{11}(u)}{4 m^2} (\ue^{2 m \gr} + 1) \log(1 + \ue^{- 2 m \gr}) \nnr
&& + \fr{f_{11}(u)^2}{16 m^2} \{ 4 m^2 \gr^2 - (\ue^{2 m \gr} + 1)
[(1 + \log (1 + \ue^{- 2 m \gr}))^2 - 1] + 2 \uL\ui_2 (- \ue^{2 m \gr}) \} \, .
\nnr
\eea


\pa{\underline{$\mu = - m$}:}


The $\mu = - m$ solution is
\bea
\ds & = & \ud \gr^2 + 2 \,  \ud u \,  \ud v - m^2 v^2 \sech^2 (m \gr) \, 
\ud u^2 - 4 m v \tanh (m \gr) \,  \ud u \,  \ud \gr \nnr
&& + [v \log(1 + \ue^{- 2 m \gr}) f_{11} (u) + f_0 (u, \gr)] \,  \ud u^2 \, ,
\eea
where
\bea
\fr{f_0 (u, \gr)}{1 + \ue^{- 2 m \gr}} & = & \gr \ue^{2 m \gr} f_{01} (u)
- \fr{\dot{f}_{11}(u)}{4 m^2} (\ue^{2 m \gr} + 1) \log(\ue^{2 m \gr} + 1) \nnr
&& - \fr{f_{11}(u)^2}{16 m^2} \{ (\ue^{2 m \gr} + 1)
[(1 + \log (\ue^{2 m \gr} + 1))^2 - 1] +
2 \ue^{2 m \gr} \uL\ui_2 (- \ue^{2 m \gr}) \} \, . \nnr
\eea


\pa{Zero cosmological constant:}


The $m \ra 0$ limit for $W_1 = - 2 m \tanh (m \gr)$ is the same as
for $W_1 = - 2 m$, which has already been considered in Section 4.2.


\pa{Positive cosmological constant:}


The positive cosmological constant analogue has already been considered in
Section 4.2 when
considering $W_1 = - 2 m \coth (m \gr)$.


\ssse{Properties}


Again, the solution has Petrov--Segre type III for $f_{11} \neq 0$,
type N for $f_{11} = 0$ and $f_{01} \neq 0$, and type O for
$f_{11} = 0$ and $f_{01} = 0$.

For the $f_{11} = 0$, $f_0 = 0$ solution, if we define
\ben
\hat{v} = u - \fr{2 \cosh^2 (m \gr)}{m^2 v} ,
\een
then the solution takes the form
\ben
\ds = \ud \gr^2 - \fr{4 \cosh^2 (m \gr)}{m^2 (u - \hat{v})^2} \,  \ud u \,  \ud
\hat{v} \, ,
\een
which can be recognised as the round AdS$_3$.  The general solution can again be
regarded as a deformation of the round AdS$_3$, taking a generalized Kerr--Schild
form with an AdS$_3$ background metric.


\se{Conclusion}


We have constructed all Kundt spacetimes that are solutions of TMG.  The general solution is reduced to a series of ordinary differential equations: (\ref{W1Kundt}), which is non-linear, and then (\ref{f1Kundt}) and (\ref{f0Kundt}), which are linear.  We have given explicitly, by solving the differential equations explicitly, those with the further specialisation that their scalar polynomial curvature invariants are all constant.  There are several families of such solutions; they fall into two broad classes, one of which has Petrov--Segre type II and is a deformation of biaxial spacelike-squashed AdS$_3$, whilst the other is of Petrov--Segre type III and is a deformation of the round AdS$_3$.  We have found that Kundt solutions of types D, N and III have constant scalar polynomial curvature invariants.

In this paper we have focussed purely on the construction of the local form of the Kundt solutions.  It is of considerable interest to investigate global structure, and in particular to examine their asymptotic behaviour at infinity.  In \cite{seztan} it is shown that solutions of the $W_1 = - 2 m$ family of the CSI type III class obey the standard Brown--Henneaux boundary conditions \cite{Brown:1986nw} for $\mu > m$, while for $\mu = m$ they obey a weaker version of those boundary conditions \cite{hematr}.  In \cite{seztan} it is also shown that the remaining CSI type III solutions do not obey either of these boundary conditions.

A further generalization would be to drop the expansion-free condition $\na_\mu k^\mu = 0$ for the geodesic null vector field.  These are known as Robinson--Trautman solutions.  The local construction of \textit{all} solutions of TMG would be completed by finding the most general such solution, but would be a considerably more formidable task.


\section*{Acknowledgements}


We thank Harvey Reall and Yoshiaki Tanii for discussions.  This research has been supported in part by DOE Grant DE-FG03-95ER40917 and NSF Grant PHY-0555575. 


\end{document}